\documentclass{ws-ijmpd}

\begin{document}
\title{Modified Averaging Processes in Cosmology and the Structured FRW model}
\author{SHAHRAM KHOSRAVI}
\address{Physics Dept., Faculty of Science, Tarbiat
Mo'alem University, Tehran, Iran.\\
School of Astronomy, IPM(Institute for Studies in Theoretical
Physics and Mathematics), P.O.Box 19395-5531, Tehran, Iran.\\
khosravi@ipm.ir}
\author{REZA MANSOURI}
\address{Department of Physics, Sharif University of
Technology, Tehran, Iran.\\
School of Astronomy, IPM(Institute for Studies in Theoretical
Physics and Mathematics), P.O.Box 19395-5531, Tehran, Iran.\\
mansouri@ipm.ir}
\author{EHSAN KOURKCHI}
\address{Department of Physics, Sharif University of
Technology, Tehran, Iran.\\
School of Astronomy, IPM(Institute for Studies in Theoretical
Physics and Mathematics), P.O.Box 19395-5531, Tehran, Iran.\\
kourkchi@physics.sharif.edu}
\maketitle
\begin{abstract}
We study the volume averaging of inhomogeneous metrics within GR
and discuss its shortcomings such as gauge dependence, singular
behavior as a result of caustics, and causality violations. To
remedy these shortcomings, we suggest some modifications to this
method. As a case study we focus on the inhomogeneous structured
FRW model based on a flat LTB metric. The effect of averaging is
then studied in terms of an effective backreaction fluid. It is
shown that, contrary to the claims in the literature, the
backreaction fluid behaves like a dark matter component, instead
of dark energy, having a density of the order of $10^{-5}$ times
the matter density, and most importantly, it is gauge dependent.

\end{abstract}

\section{Introduction}

The amazing success of the FRW model of the universe has for years
overshadowed the fact that we have devised it for a smoothed out
geometry of the actual universe which shows inhomogeneity at
different scales. The dark energy concept cannot be properly
understood until the effect of inhomogeneities in the
observational parameters and the role of geometrical averaging is
understood. While studies of inhomogeneous models are progressing
(see Refs.~\refcite{Mansouri05},~\refcite{Celer07} for extensive
literature review, and also
Refs.~\refcite{Chuang}--~\refcite{Kolb07}), the question of how to
average the geometry is still an open question. Is it possible to
write the Einstein equations for an inhomogeneous universe, make
then an averaging of the geometry, and obtain an effective FRW
model of certain type? What would then be the difference between
the result of averaging and a $\Lambda CDM$ model of the universe
\cite{Ellis84,Ellis87}?

There have been different attempts to answer these questions. The
inhomogeneities maybe considered as a perturbation to the FRW
models; an averaging process then leads to backreaction and a
modified FRW universe
\cite{Celer07,Bild91,Futa93,Kolb05,Romano,Singh1,Singh2}. The
perturbative approach is likely to diverge due to the growth of
perturbation at the epoch of structure formation, i.e. exactly the
epoch of interest related to dark energy, as has been shown in
\cite{Notari05} (for a review of different approaches see
Ref.~\refcite{Celer07}). In the non-perturbative approach
\cite{Zalal92,Zalal93,Buchert00,Buchert01} a spatial volume
averaging is devised to smooth out the inhomogeneities of the
geometry as well as the fluid content of the universe leading to a
non-standard homogeneous FRW model. Although the methods are
non-perturbative, the difference between the real universe and the
averaged one is usually called backreaction too.

In this paper we are dealing with the Buchert's non-perturbative
approach to the averaging problem in general relativity, which is
based on the averaging formalism within the Newtonian gravity
\cite{Buch-Ehl-1}. While the Newtonian case, being
non-relativistic, is well defined, the general relativistic case
is to be applied with care \cite{Ishi06}. The main critique to the
non-perturbative procedures is the fact that any inhomogeneous
cosmological solution leads quickly to singularities making the
formalism invalid \cite{Ishi06}. Using flat LTB cosmological
models, we will see in detail how the singularities affect the
volume averaging procedure. Based on resulting insights, we use
the structured FRW model of the universe (SFRW) \cite{Khosr07} to
propose and apply two different modifications of the volume
averaging methods along the lightcone. It is then shown, that the
averaging and the resulting backreaction is coordinate dependent,
corresponds to dark matter instead of dark energy, and its
corresponding density is 4 to 5 order of magnitudes less than the
mean density of the universe.

Section 2 is devoted to flat LTB metrics, a short introduction to
the structured FRW model of the universe, and a discussion on the
place of the past light cone and the SFRW singularities within it.
After introducing the volume averaging method within GR in section
3, we continue with a critique of the Buchert's averaging method
elaborating its shortcomings, suggest some modifications to it,
and calculate the backreaction in SFRW model for different
methods. We conclude then in section 4.

\section{The structured FRW model of the universe}

\subsection{Overview}

\label{sec-Flat}  The structured FRW (SFRW) model proposed
recently \cite{Khosr07} is a suitable case to study different
averaging methods in general relativistic cosmological models. The
basic idea in developing the SFRW model is the mere fact that the
events outside the past lightcone of the observer have had no
influence on the past events observable to us. One may then ignore
the possible inhomogeneities outside the lightcone and model the
universe outside the lightcone as a FRW one. It then has to be
matched to an inhomogeneous universe inside the past lightcone. As
the structures are effectively influencing the universe much later
than the last scattering surface, SFRW model is applied to the
matter dominated and pressure-less universe, and up to those
$z$-values corresponding to the length scales of at least 1000 Mpc
in which the universe is almost homogeneous. In contrast to the
familiar usage of LTB metric to represent over- or under-density
bubbles in the constant time slices of the universe, in SFRW the
LTB junction to FRW is adapted along the past lightcone.

\begin{figure}[h]
\centering
\includegraphics[angle=0, scale=.5]{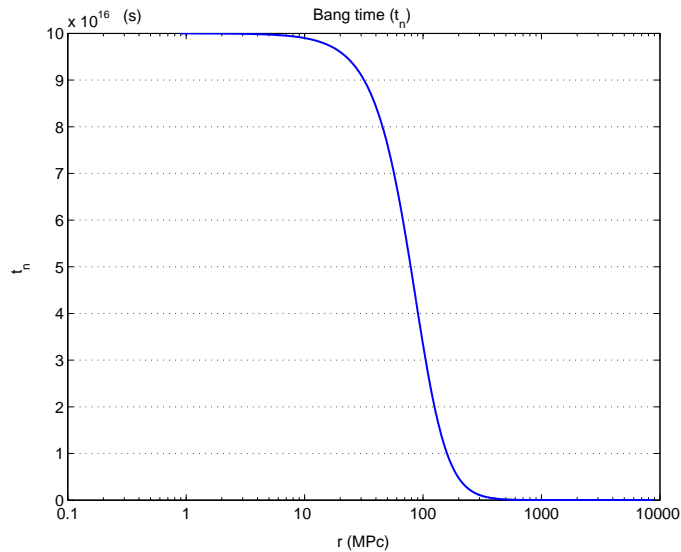}
\caption{ $t_n$ vs r for $t_{n}=\alpha/(r^4+r^2+1)$.} \label{fig1}
\end{figure}

Although, FRW and LTB could in principle be any of the three cases
of open, flat, or close, it has been shown in
Ref.~\refcite{khakshour07} that the only meaningful matching of
these two spaces along a null boundary is the flat-flat case.
Given the cosmological preferences, we are therefore assuming both
metrics to be flat. As a consequence of this matching along the
past lightcone, the observed density of the universe along the
past lightcone has to be set equal to the density of FRW metric
everywhere outside the lightcone \cite{khakshour07}. The
inhomogeneities are therefore just in our neighbourhood within the
lightcone up to distances of the order of 1000 Mpc.

 We have to stress however that SFRW is just a toy model to understand the
impact of nearby inhomogeneities on our observations inside our
past lightcone; one is free to use it as a base for a Swiss cheese
model, or add extra perturbations to it. It should not be
considered as a model for structure formation similar to the onion
model \cite{Notari05}.

\subsection{The Flat LTB solution}

\label{sec-bang} According to the SFRW model, a flat FRW universe
outside the past lightcone is matched to a marginally bound or
flat LTB metric inside the lightcone. Therefore, we restrict
ourselves in this paper to the marginally bound LTB case. These
are solutions of Einstein equations described by the metric
\begin{eqnarray}\label{metric}
ds^{2} = -c^2 dt^{2} + R^{'^{\,2}}\,dr^{2} +
R^{2}(r,t)(d\theta^{2}+\sin^{2} \theta d\phi^{2}),
\end{eqnarray}
in which overdot and prime (will thereafter) denote partial
differentiation with respect to $t$ and $r$, respectively. The
corresponding Einstein equations turn out to be
\begin{eqnarray}\label{field}
\dot{R}^{^{2}}(r,t) &=& \frac{2GM(r)}{R} ,\\
4\pi\rho(r,t) &=& \frac{M'(r)}{R^{^{2}} R^{{\,'}}}.
\end{eqnarray}
The density $\rho (r,t)$ is in general an arbitrary function of
$r$ and $t$, and the integration time-independent function $M(r)$
is defined as
\begin{equation}\label{density}
M(r)\equiv 4\pi \int^{R(r,t)}_{0}\rho(r,t)R^{\,2}dR
     = \frac{4 \pi}{3} \bar{\rho}(r,t) R^{\,3},
\end{equation}
where $\bar{\rho}$, as a function of $r$ and $t$, is the average
density up to the radius $R(r,t)$. The metric (\ref{metric}) can
also be written in a form similar to the Robertson-Walker metric.
The following definition
\begin{equation}
a(t, r) \equiv \frac{R(t, r)}{r}\,\, ,
\end{equation}
brings the metric into the form
\begin{equation}
ds^2 = -c^{\,2}dt^2 + a^2\left[\left(1 + \frac{a' r}{a}\right)^2
dr^2+ r^2 d\Omega^2 \right].
\end{equation}
For a homogeneous universe, $a$ doesn't depend on $r$ and we get
the familiar Robertson-Walker metric. The corresponding field
equations can be written in the following familiar form
\begin{equation}
\big (\frac{\dot a}{a} \big)^2 =  {1\over 3}\frac{\rho_c(r)}{a^3},
\end{equation}
where we have introduced $\rho_c(r) \equiv 6M(r)/r^3$ indicating a
quasi comoving $r$-dependent density. These are very similar to
the familiar Friedman equations, except for the $r$-dependence of
different quantities. The solutions to the field equations can be
written in the form
\begin{eqnarray}
R(r,t) &=&  \left[\frac{9GM(r)}{2}\right]^{1\over 3}
\left[t- t_n(r)\right]^{2\over3}, \nonumber  \\
a(r,t) &=& \left[\frac{3}{4}\, G\rho_c(r)\right]^{1\over
3}(t-t_n(r))^{2\over3}.
\end{eqnarray}
These are by now standard definitions of the flat LTB metric as a
solution of Einstein equations \cite{Khosr07}.

 Furthermore, the expansion and shear are defined in the following way:
\begin{eqnarray}
\Theta_{ij} &=& - K_{ij} = h^{\mu}_i h^{\nu}_j u_{\mu;\nu},\\
h_{\alpha \beta} &=& g_{\alpha \beta} +u_{\alpha} u_{\beta}\,\, ;
\hspace{0.5cm} h_{ij} = g_{ij},
\end{eqnarray}
For the LTB metric (\ref{metric}) we obtain

\begin{eqnarray}
\Theta^{i}_{j} &=& \left[\frac{\dot R'}{R'}\,\, , \frac{\dot R}{R}\,\, , \frac{\dot R}{R}\right],\\
\theta &=& tr \Theta^{i}_{j} = \frac{\dot R'}{R'} + 2 \frac{\dot R}{R},\\
\sigma^{i}_{j} &=& \Theta^{i}_{j} - {1\over 3}\theta \delta^{i}_{j},\\
\sigma^{1}_{1} &=& -2 \sigma^{2}_{2},\\
\sigma^{2}_{2} &=& \sigma^{3}_{3} = {1\over3}\left(\frac{\dot R}{R} - \frac{\dot R'}{R'}\right),\\
\sigma^2 &=& {1\over3} \left(\frac{\dot R}{R} - \frac{\dot
R'}{R'}\right)^2.
\end{eqnarray}

\begin{figure}
\centering
\includegraphics[angle=0, scale=.5]{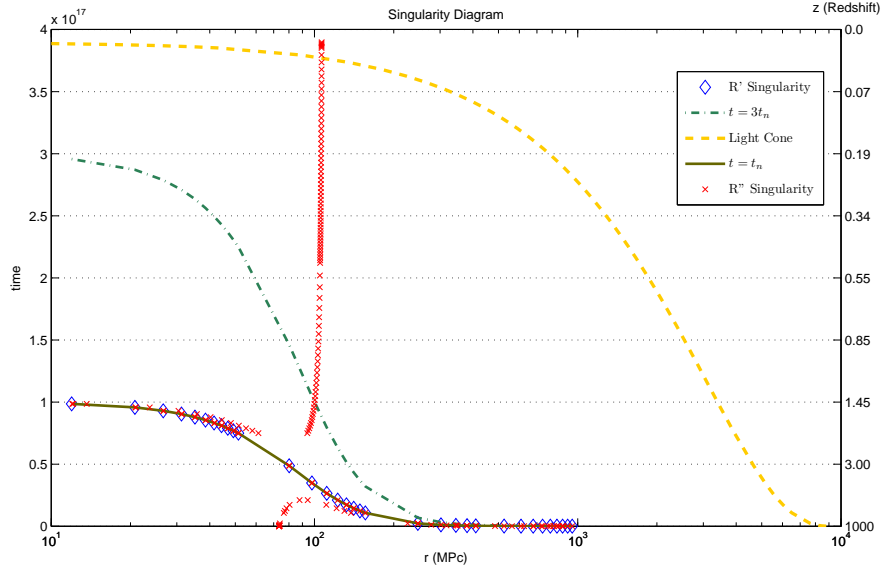}
\caption{ Lightcone (time vs $log_{10}(r)$) for
$t_{n}=\alpha/(r^4+r^2+1)\,\, , \alpha=10^{17} sec$. The solid
line curve shows the shell focusing area $R = 0$ at $t=t_n$. On
this curve both $R'$ and $R''$ are singular. All singularities are
well inside the curve $t=3t_{n}(r)$. The $\dot{R^{\prime}}$
singularity (corresponding to
$t=t_{n}-\frac{1}{3}rt_{n}^{\prime}$) are not shown because they
almost lie completely on the curve for $t=t_{n}(r)$. The
dash-dotted curve corresponds to Kretschman singularity
($t=3t_{n}(r)$). The curve plotted with cross signs corresponds to
zero of $R''$; None of the geometrical scalars become singular on
this curve.} \label{fig2}
\end{figure}

Now, the flat LTB part of the SFRW model is defined by its bang
time which is assumed to be
\begin{equation}\label{bang}
t_n = \frac{\alpha}{r^4+r^2 + 1},
\end{equation}
where $r$ is scaled to 100 Mpc, as a typical inhomogeneity scale.
For $r$-values much larger than 100 Mpc the bang function $t_n$
tends to zero and, therefore, the LTB tends to a FRW metric. The
special form of the bang function has been chosen such that there
should be no singularity on the past lightcone. The constant
$\alpha=10^{17} sec$ is chosen such that the age of the universe
at the last scattering surface corresponds to that of the standard
cosmological model \cite{Khosr07}.
 The comoving coordinate $r$ is
used alternatively in the scaled form or not, and the reader may
simply see from the context which one is meant. Figure \ref{fig1}
shows $t_n$ as a function of $r$. The bang time is almost
everywhere zero except in our vicinity, reflecting the desired
feature of SFRW. It can also be seen that for large $r$
corresponding to the redshifts $z>1$, we have almost FRW.
 Now, for this bang time we have
\begin{equation}
t'_n|_{r = 0} = 0. \label{weak}
\end{equation}

Therefore, there is no weak singularity at the origin
\cite{Vander06}. In fact, for the LTB domain with this bang time,
there is no singularity in the vicinity of the lightcone, as it is
shown in figure \ref{fig2}. No invariant of the metric has a
singularity within the domain of our interest. All of the
quantities appearing in the averaging process behave regularly at
$t = 3t_n$ where the Kretschmann invariant is singular. The
singularities of LTB metric, which are more sophisticated than in
the case of the Robertson-Walker metric, have been discussed
extensively in the literature (see for example
Ref.~\refcite{Vander06}-- ~\refcite{Newman86}). Vanishing of each
of the metric functions and its derivatives $R, R', \dot R, \ddot
R, \dot R^{\,'}, R^{\,''}$ may lead to different singularities. In
a general LTB metric there is another singularity, the event
horizon, related to zero of $1+ E$, where $E(r)$ is the energy
function of the LTB metric absent in our flat LTB case.
The place of different singularities are summarized as follows:
\begin{eqnarray}
& &R^{\,\prime},R^{\,\prime\prime},\dot{R},\dot{R^{\,\prime}} = \infty \hspace{.35cm}\longrightarrow\hspace{.5cm} t=t_n\,\, , \nonumber\\
& &R =0 \hspace{2.6cm}\longrightarrow\hspace{.5cm} t=t_n\,\, ,  \nonumber\\
& &R^{\,'} = 0 \hspace{2.5cm}\longrightarrow\hspace{.5cm} t=t_n+\frac{2}{3}rt'_n\,\, ,\\
& &\dot{R^{\,'}}= 0 \hspace{2.5cm}\longrightarrow\hspace{.5cm} t=t_n-\frac{1}{3}rt'_n\,\, , \nonumber\\
& &R{\,''} = 0 \hspace{2.45cm}\longrightarrow\hspace{.5cm}
t=\frac{3rt_nt''_n+6t_nt'_n -r{t'_n}^2}{6t'_n+3rt''_n}\nonumber.
\end{eqnarray}

It is obvious that for $r\ll 1$, the bang time approaches a
constant, and for $r\gg 1$ it approaches a constant, in fact zero,
meaning that for large $r$ we have effectively FRW metric again.

\begin{figure}
\centering
\includegraphics[angle=0, scale=.6]{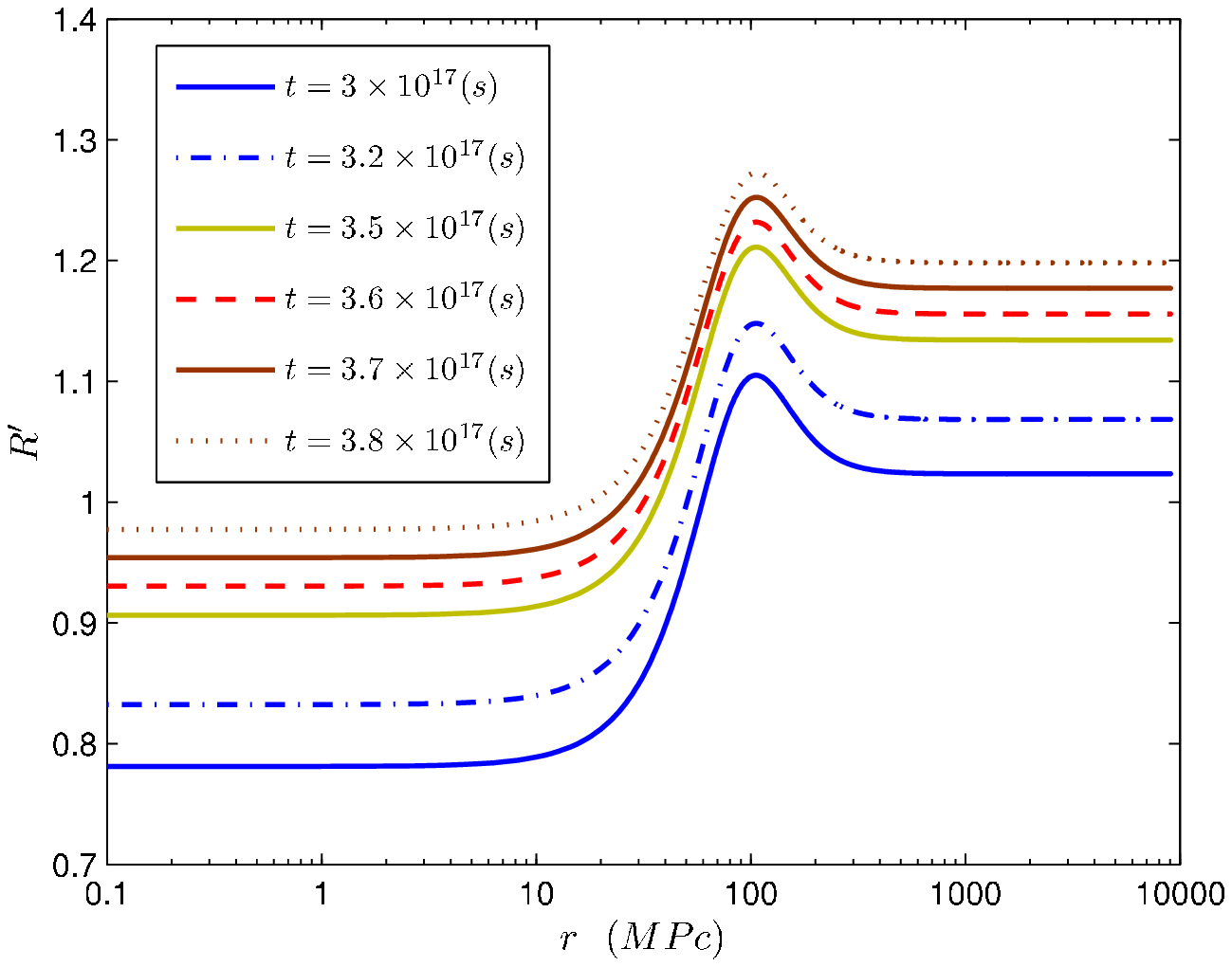}
\caption{ $R^{\,'}$ is plotted versus $r$ for different time
values. Looking at $R{\,'}$ as an effective scale factor, it shows
that the scale of the universe increases with time, although the
rate of cosmic expansion is different at different places. }
\label{fig3}
\end{figure}
 The zero's of $R^{\,''}$ are also sketched in the
mentioned figures. The corresponding curve intersects the past
lightcone on a point which corresponds to a local maximum of
$R{\,'}$ as can be seen from figure \ref{fig3}, and none of the
metric invariants, including that of Kretschman, have a
singularity at this event. Therefore, $R^{\,'' }= 0$ does not
correspond to a metric singularity.

Having established the fact that singularities disturbing the
averaging process happen around $t=t_n$, we note that the time
constant slices with no singularities corresponds to redshift
values less than $z = 1.45$. Therefore, any averaging over a fixed
domain includes singularities except for $z$-values less than
1.45. Note that even in domains of no singularity, where the
averaged values may be defined, the domain gets extended beyond
the lightcone and we have to face superluminal effects. Finally,
we stress again that for this special choice of the bang time the
condition (\ref{weak}) is valid and therefore, there is no weak
singularity at the origin \cite{Vander06}.

\section{ Averaging in cosmology}

\subsection{Volume averaging in GR}

The cosmic fluid is assumed to be perfect and irrotational. A
flow-orthogonal foliation of spacetime, i.e. synchronous slicing,
with the metric $ds^2 = -dt^2 + g_{ij}dx^i dx^j\,$ is then
assumed. We should note that the synchronous coordinates, being
not necessary for the following definitions, are however suitable
for our problem. One could equally choose Newtonian gauge which is
more suitable to study backreaction of inhomogeneities considered
as perturbation to FRW models.

 The volume-averaging is based on the following definition. Let
$f(\vec{r}, t)$ be an arbitrary function of coordinates. Its
average is defined by
\begin{equation}
\langle f\rangle \equiv {1 \over V_D} \int_D dV f,
\end{equation}
where $dV$ is the proper volume element of the 3-dimensional
inhomogeneous domain $D$ we are considering and $V_D$ is its
volume. The space-volume average of the function $f(\vec{r}, t)$
does not commute with its time derivative:
\begin{equation}\label{nc}
\langle f\rangle^{\cdot} - \langle \dot f\rangle = \langle
f\theta\rangle - \langle f\rangle \langle \theta \rangle,
\end{equation}
where the expansion scalar $\theta$, being equal to the minus of
the trace of the second fundamental form of the hypersurface $t =
const.$, is now a function of $r$ and $t$. The right hand side
trivially vanishes for a FRW universe because of the homogeneity.
The averaged scale factor is then defined using the volume of our
domain $D$ by $a_D \equiv V_{D}(t)^{1 \over 3}$. Now it can be
shown that \cite{Buch-Ehl-1}
\begin{equation}\label{avtheta}
\theta_D \equiv \langle \theta\rangle \equiv  {\dot V \over V} =
3{\dot a_D \over a_D} = 3 H_D,
\end{equation}
where we have used the notation $\dot a_D \equiv {da_D/dt}$, and
denoted the average Hubble function as $H_D$. Therefore, all the
derived quantities should be based on the average value $a_D$.
This is why one usually takes the above definition for the mean
Hubble parameter and not $\langle\dot a/a\rangle$, which is
different from ${\dot a}_D/a_D$. A similar difference holds for
the second derivative of $a$:
\begin{equation}
 \left<\frac{\ddot a}{a} \right>\neq \frac{\langle \ddot a \rangle}
{\langle a \rangle} \neq {\ddot a_D \over a_D}.
\end{equation}
Therefore, the definition of the averaged deceleration parameter
is not without ambiguity, specially because there is no firm
relation like (\ref{avtheta}) for the deceleration parameter. This
has motivated many authors so far to make the following definition
for the deceleration parameter:
\begin{equation}
q_D  = - \frac{{\ddot a}_D a_D}{{\dot a_D}^2} =  - \frac{{\ddot
a}_D}{a_D}\frac{1}{H_D^2}.
\end{equation}
Now, the averages of the Einstein equations, using the Hamiltonian
constraint and the Raychaudhuri equation, is written in the
following form \cite{Buchert00}:
\begin{eqnarray}\label{ray1}
\left({\dot a_D \over a_D} \right)^2 = {8\pi G\over3} (\rho_D +
\rho_{\Sigma} +\rho_{\cal R}), \\
\label{ray2}\frac{\ddot a_D}{a_D} = -{4\pi G\over3}(\rho_D
+4\rho_{\Sigma}),
\end{eqnarray}
where we have set $\langle \rho \rangle = \rho_D$, $\rho_{\Sigma}
=
 \Sigma/8\pi G$, $\rho_{\cal R} = -{\cal R}_D/16\pi G$,
 $\cal R_D$ is the backreaction term due to the average of three
dimensional Ricci scalar, and $\Sigma$ is the backreaction term
corresponding to the non-vanishing shear defined as follows:
\begin{equation}
\Sigma \equiv \langle \sigma^2 \rangle - {1\over3}\left< (\theta -
\langle \theta \rangle)^2 \right> = \langle \sigma^2 \rangle -
{1\over3}\left[\langle \theta^2\rangle - \theta_D^2\right],
\end{equation}
where $\sigma$ is the shear scalar. It may be written more
suitably in the form
\begin{eqnarray}\label{sigma-A}
\Sigma &=& \langle \sigma^2 \rangle - {1\over3}[\langle
\theta^2\rangle -\theta_D^2] = \left[\langle \sigma^2 \rangle -
{1\over3}\langle \theta^2\rangle\right]
+ {1\over3}\theta_D^2\nonumber\\
&=& \left[\langle \sigma^2 \rangle - {1\over3}\langle
\theta^2\rangle\right] +3H^{2}_D = A +3H^{2}_D,
\end{eqnarray}
where
\begin{equation}\label{A}
A \equiv \left[\langle \sigma^2 \rangle - {1\over3}\langle
\theta^2\rangle\right].
\end{equation}
Substituting from (12)-(16) we obtain
\begin{equation}
A = \left<\left[ -2\frac{\dot R' \dot R}{R' R}- \frac{\dot
R^2}{R^2}\right] \right> .
\end{equation}
Note that in the synchronous gauge we have chosen, the so-called
dynamical backreaction term vanishes and $\Sigma$, the kinematical
backreaction, is equal to the total backreaction
\cite{Buchert01,Behrend}. We use the backreaction here for the
geometric term $ \langle G_{\mu\nu}\rangle - G_{\mu\nu}$, where
$G_{\mu\nu}$ is the Einstein tensor and the averaging is
understood to smooth out the inhomogeneities. Therefore, the
Einstein tensor $G_{\mu\nu}$ is divided into a ``homogeneous" part
corresponding to the FRW universe, and a remaining part which
could be brought to the rhs of the Einstein equations as
backreaction of the inhomogeneities and interpreted as the
energy-momentum tensor of a ``geometric fluid". Furthermore, the
total density $\rho$, and the backreaction density $\rho_b$, may
be defined in the following way:
\begin{eqnarray}
\rho &=& \rho_D + \rho_b,\\
\rho_b &=& \rho_{\Sigma} + \rho_{\cal R}=\frac{-1}{8\pi
G}\left(-\Sigma + {1\over2}\cal R_D\right).
\end{eqnarray}
According to this definition, the field equations lead to the
following effective backreaction pressure:
\begin{equation}\label{state}
p_b = \rho_{\Sigma} - {1\over3}\rho_{\cal R}.
\end{equation}
Although $\theta_D$ and $H_D$ are proportional, $\langle \theta^2
\rangle$ and $\langle H^2 \rangle$ are not. Hence, (\ref{sigma-A})
and (\ref{A}) cannot be written in terms of $H$. In the case of
vanishing the three-dimensional Ricci scalar, as it is the case in
the flat LTB, the equation of state reduces to
\begin{equation}
\rho_b = \rho_{\Sigma}\, \,; \ \ \ p_b = \rho_{\Sigma}.
\end{equation}
Therefore, in the flat LTB case the backreaction is defined
effectively by an ideal fluid having the equation of state
\begin{equation}
\rho_b = p_b = \rho_{\Sigma} \,\,\, ;\,\,\, w = 1.
\end{equation}
It is clear that it can not be interpreted as dark energy. We have
not verified the gauge dependence of the above equation of state,
as it is beyond the scope of this paper. It may be different in
Newtonian gauge, which would be another sign of coordinate
dependence of the averaging method and its consequence for the
backreaction.

\subsection{Critique of the averaging method and modifying
suggestions}

The averaging process so far described and used in literature has
already been criticized in Ref.~\refcite{Ishi06}. Here we review
different
shortcomings of the method: \\
i) It is generally gauge dependent. One then has to check each
time how far the result in a specific gauge is viable. Note that
the Eqs. (26-27) are the result of a non-covariant integration
over the Raychaudhuri equation. This has not to be confused with
the assumption of homogeneity of the universe and the resulting
FRW universe: the FRW universe {\it is} a solution of the Einstein
equations
based on the homogeneity assumption. \\
ii) Even if there is some natural choice of coordinates, like the
comoving synchronous gauge, i.e. slices orthogonal to the world
lines of the dust, it usually breaks down due to the formation
of caustics \cite{Ishi06}. \\
iii) The domain of integration over which the average values are
defined is fixed. In cosmological models of interest, even in
regions without any caustics, the domain $D$ extends definitely
outside the lightcone for the most interesting range of time or
redshift values, independent of how small the chosen range of $r$
is. Integrating over distances outside the lightcone is, however,
equivalent to taking into account superluminal effects. This is
clearly a noncausal procedure which should not be implemented in
the relativistic equations. It is equivalent to assigning a value
to the density at a point on the light cone, or within it,
depending on events causally disconnected to it, similar to the
horizon problem in standard cosmology.

\subsubsection{Gauge dependence}
According to the Stewart-Walker lemma \cite{Stewart}, any average
quantity is gauge invariant if the zeroth order part vanishes on
the background, or is a constant scalar field there, or is a
linear combination of products of Kronecker deltas with constant
coefficients. The theorem has been used in Ref.~\refcite{Schwarz}
to show the gauge invariance of the quantities they are
calculating. One should however be aware of the fact that the
theorem is valid only for a perturbed metric with respect to a
background. Therefore, this theorem can not be applied to the
general case of an exact solution of Einstein equations such as
the LTB metric we are using here. In fact, it is not clear at all
if and when the LTB metric can be written as a perturbation to FRW
metric for cases where backreaction maybe significant
\cite{Ghassemi}. We will show in the section \ref{modified} that
the backreaction is non-zero for the structured FRW model in the
gauge chosen, although it is exactly zero in another gauge
\cite{Singh1}, indicating the gauge dependence of the averaging
method.

\subsubsection{Modified methods: Lightcone averaging formalism}

\label{modified}

We suggest two different alternatives to the volume averaging
method in general relativity \cite{Buchert00} to the aim of
circumventing the shortcomings of the method elaborated in the
section (3.2). The general motivation is to remedy the non-causal
input to the averaging integrals through the fixed domain and also
to avoid singularities. The first is accomplished by a time
constant integral up to the lightcone. To avoid the singularities,
we consider a region bounded by the lightcone and a time-like
hypersurface within it at a distance of the order of 1000 Mpc to
be sure that outside this inhomogeneous domain we may consider the
universe as homogeneous again. Note that the distance from any
point on the curve corresponding to $t = 3t_n$, i.e. the
Kretschmann singular curve, is at least of the inhomogeneity order
of 1000 Mpc (see fig \ref{fig2}). We may therefore assume that the
domain within the lightcone but outside the Kretschmann curve is
effectively homogeneous. We have already mentioned that the
density for the events on the Kretschmann curve are regular.
Therefore, these events are no obstacle to define
the averaging if necessary.\\
Both modified domains are bounded by the lightcone. To distinguish
them we call the averaging by using the first domain the
in-lightcone and the second one the on-lightcone averaging method.
Let us summarize the three different averaging procedures:
 1) {\it Volume averaging using fixed domain}\\
This is the standard averaging formalism used in literature
\cite{Buchert00}.

 2) {\it In-lightcone averaging formalism}\\
The domain of integration is extended from a fixed $r$-value, say
$r = 0$, to the corresponding $r$ at the lightcone for a fixed
time. Clearly the non-causal character of the standard volume
averaging is absent here, but the domain may still includes
singular points.
 From what has been outlined in sections \ref{sec-Flat} and
\ref{sec-bang}, both the volume averaging with fixed domain and
the in-lightcone averaging method may include caustics. Hence,
these methods are only applicable to small $z$ values where no
singularity is seen within the domain of integration. In our model
we are bounded to the redshifts $z < 1.45$.

 3) {\it On-lightcone averaging formalism}\\
A time-like hypersurface within the lightcone, far from the
singularities but at a distance from the lightcone of the order of
the inhomogeneity scale such as the Kretschmann curve is chosen.
The domain of integration is then defined for a constant time from
a point on this hypersurface up to the lightcone. By this
on-lightcone formalism we have secured the causal implications of
the averaging
integrals and also a singularity-free domain of integration.\\

The first two methods are applicable along the past lightcone just
up to z-values where the density is regular for any $r$ value
corresponding to $z \approx 1.45$. As it is seen from the figure
\ref{fig2}, all the metric singularities lie under the curve
corresponding to $t=t_{n}(r)$. Therefore, the hypersurface
necessary for the on-lightcone method may be any time-like one
within the lightcone bounded by the curve $t = t_n$ such as the
Kretschmann curve. The averaging integrals in this case are
well-defined and maybe handled numerically.

  The on-lightcone averaging method is, therefore, free from both
shortcomings and we may define all along the lightcone without
incorporating any caustics or non-causal effects.

\subsection{Results}
\begin{figure}[t]
\centering
\includegraphics[angle=0, scale=.55]{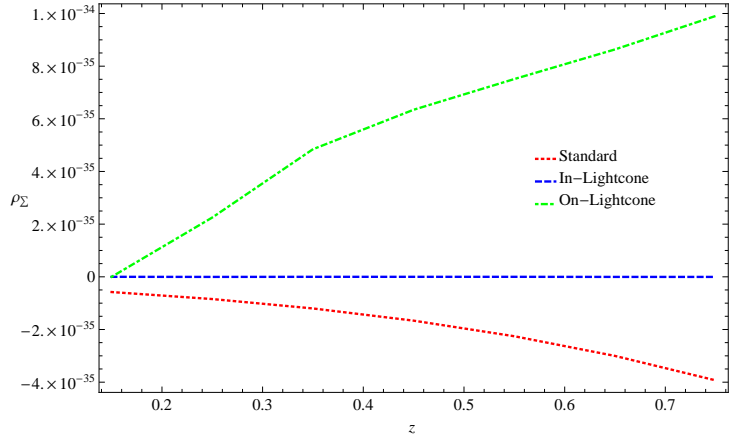}
\caption{Comparison of the backreaction density as a function of
redshift by the three averaging methods for low redshifts. The
value of $\rho_\Sigma$ is not defined for redshifts higher than
$\sim1$ due to the singularities in standard and In-lightcone
methods.} \label{fig4}
\end{figure}
 Now, we have all the prerequisites to do the
averaging. The results are plotted in the form of backreaction
density $\rho_{\Sigma}$ and the ratio of backreaction to the
matter density $\rho_{\Sigma}/\rho_{z}$ versus redshift for the
bang time ({\ref{bang}) (figures \ref{fig4}-\ref{fig6}).
\begin{figure}[t]
\centering
\includegraphics[angle=0, scale=.6]{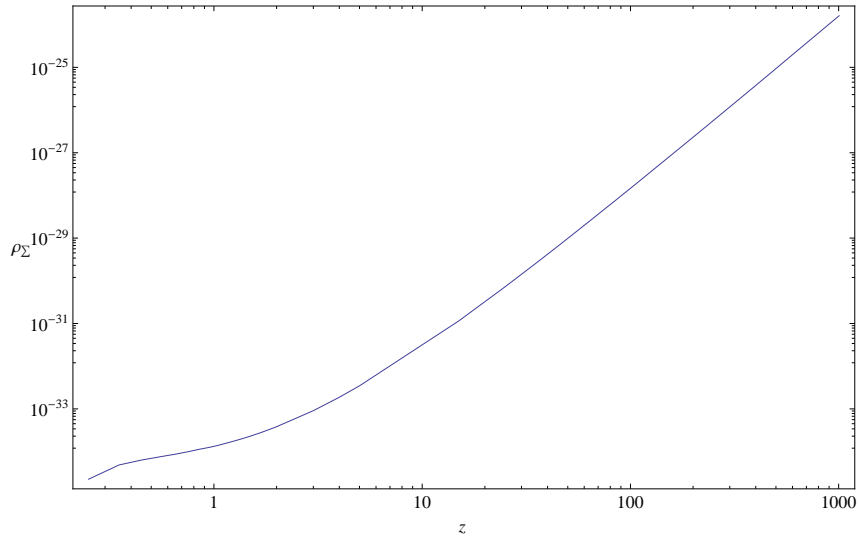}
\caption{Backreaction density as a function of redshift by the
on-lightcone averaging. } \label{fig5}
\end{figure}
\begin{figure}
\centering
\includegraphics[angle=0, scale=.55]{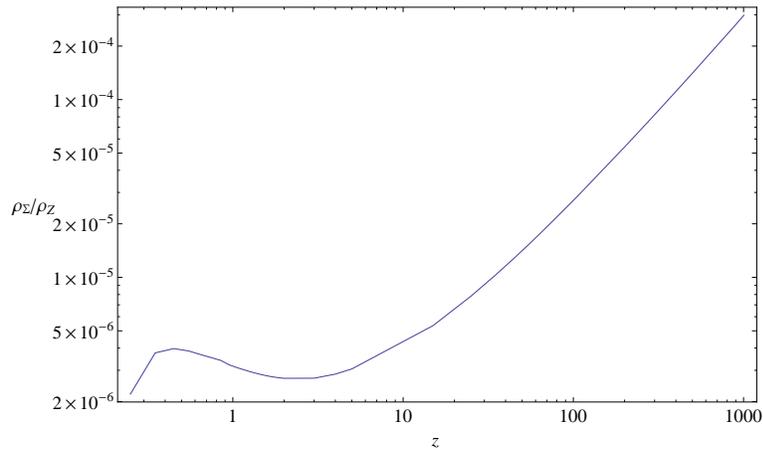}
\caption{Ratio of the backreaction density to the matter density
by the on-lightcone averaging.} \label{fig6}
\end{figure}
 Figure 4
shows the backreactions up to $z \approx 1.5)$ for the three
methods. Larger $z$-values leads to singularities and, therefore,
the averaging cannot be executed in the volume averaging and
in-lightcone methods. This shows again the restricted
applicability of the volume averaging method. In the case of
on-light-cone method we have been able to integrate up to the
surface of the last scattering corresponding to $z \approx 1100$.
Therefore, any meaningful statement about the effect of
backreaction can only be made for this method. For the sake of
completeness, and as a reference to our numerical calculation, we
have also calculated numerically the case of bang time $t_n=0$
equivalent to FRW with vanishing backreaction plotted in ( figures
\ref{fig7}, \ref{fig8}). The vanishing of backreaction in this
case is easily seen up to the numerical errors.
 In all the cases
under consideration, the backreaction density and pressure are
positive and roughly $5$ order of magnitudes smaller than the dark
energy needed to explain the acceleration of the universe. Hence,
it cannot be concluded that the backreaction has any effect
towards explaining the mysterious dark energy. The backreaction
for a LTB universe using the volume averaging for a fixed domain
(figure \ref{fig4}) shows clearly that the effect is even three
order of magnitudes smaller than all the other values for the SFRW
model.
 A noticeable effect is the
increase of the backreaction with redshift after its minimum at
about $z \approx 4$. This is in contrast to the motivation of SFRW
in choosing the bang time (\ref{bang}). One would expect that the
backreaction would decrease all along the past lightcone. To
understand this behavior, we have plotted the bang time
(\ref{bang}) versus the redshift in figure 9. It shows clearly
that the bang time has a minimum at roughly the same redshift
value of the minimum of the backreaction. This redshift value
corresponds to $r \approx$ 5000 Mpc which is of the order of
magnitude of the inhomogeneity of the universe. It is, therefore,
clear why the backreaction increases after that minimum: the bang
time, and as a consequence the deviation of LTB from FRW,
increases. The selected bang time (\ref{bang}) seems to be
well-behaved up to the inhomogeneity scale. We know already that
the universe is homogeneous at that scale to a very good
approximation, and is modeled by FRW standard cosmology. Hence, we
may assume that within the SFRW philosophy the metric after $z
\approx 4$ is given by FRW.\\
 After publishing
the first version of this work, a study \cite{Kolb07} appeared,
discussing an averaging approach along the past lightcone, in
contrast to time constant domains. The authors define a new
procedure for averaging in cosmology not related to the volume
averaging and its modifications we have proposed in this paper. A
separate study on the backreaction in a flat LTB model universe
\cite{Mattsson} was published, in which the authors suggest a
so-called 'running averaging scale' to modify the volume averaging
method, using a gauge similar to that of Ref.~\refcite{Singh1}
and, as in Ref.~\refcite{Singh1}, show the vanishing of the
backreaction in it. They are mainly interested in the modification
of the luminosity distance. A recent paper on the cosmological
backreaction from perturbations \cite{Behrend} also appeared.
Working in the Newtonian gauge, the authors announce a
backreaction of the order of $10^{-5}$ times the matter density,
similar to our result, but with an effective equation of state
$w\approx -1/19$.

\section{Conclusion}

\begin{figure}[t]
\centering
\includegraphics[angle=0, scale=.55]{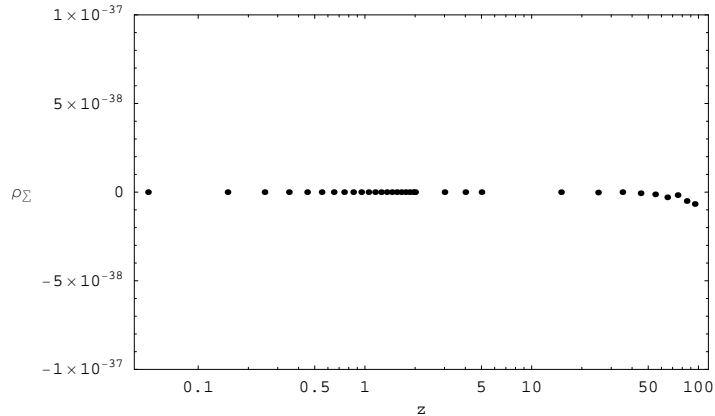}
\caption{Backreaction density as a function of redshift by the
on-lightcone averaging for $t_n=0$. } \label{fig7}
\end{figure}

\begin{figure}
\centering
\includegraphics[angle=0, scale=.6]{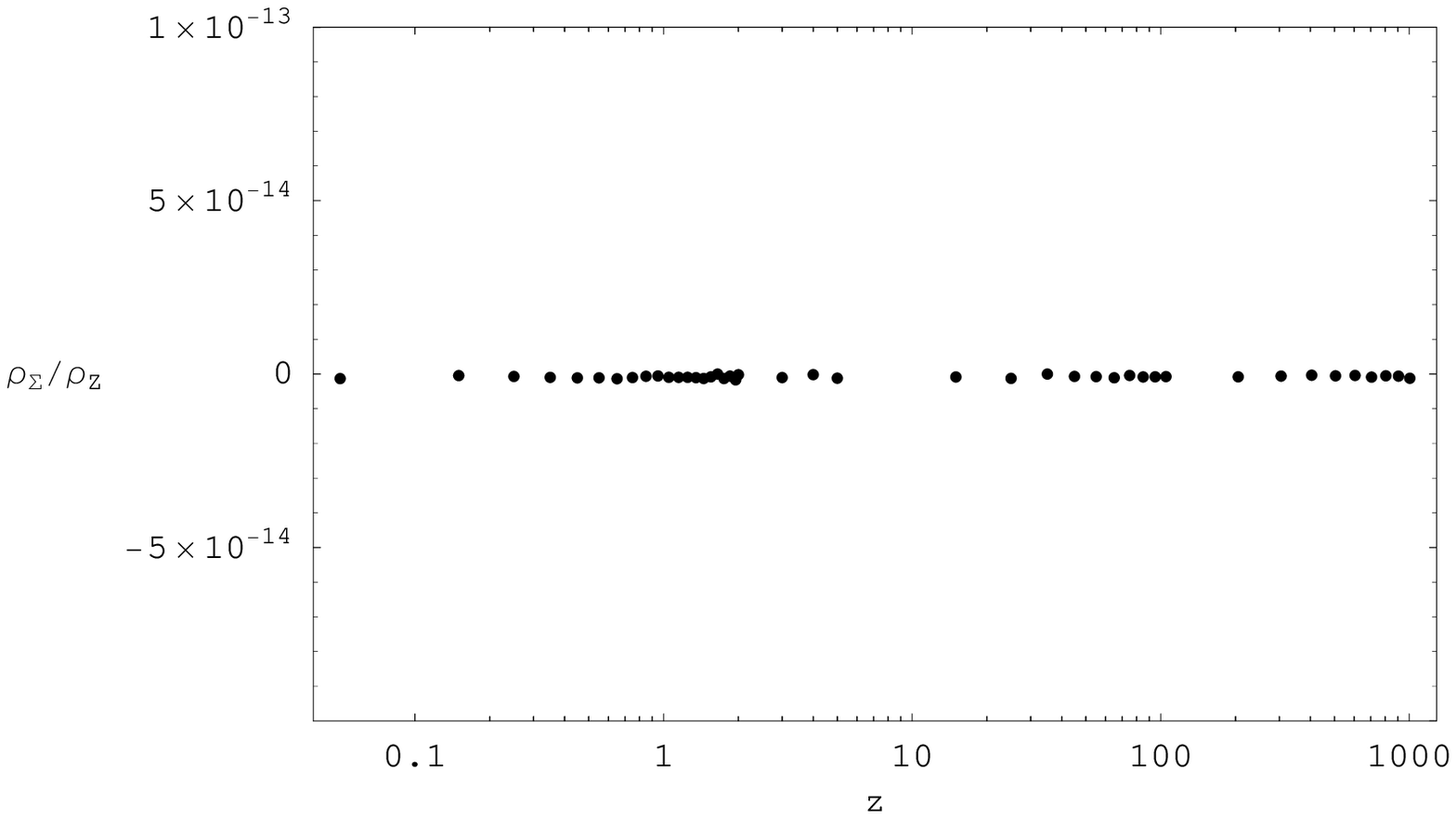}
\caption{Ratio of the backreaction density to the matter density
by the on-lightcone averaging for $t_n=0$.} \label{fig8}
\end{figure}

\begin{figure}
\centering
\includegraphics[angle=0, scale=.6]{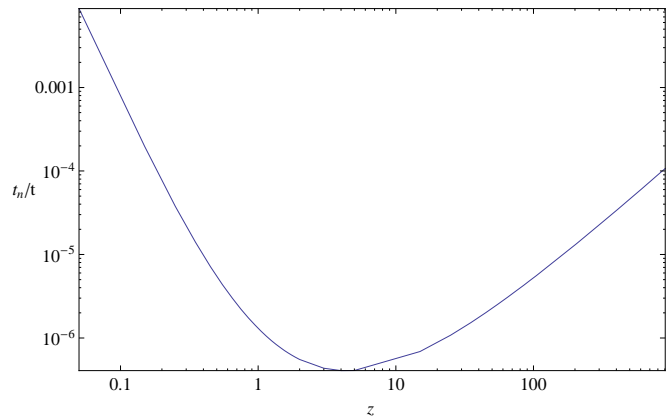}
\caption{Ratio of the bang time to the time on the lightcone
versus redshift.} \label{fig9}
\end{figure}

We have explored the possibility of explaining, at least
partially, the dark energy using the volume averaging of Einstein
equations for a specific model based on the flat LTB inhomogeneous
solutions. We were able to define two modifications to the
familiar method with a fixed domain of integration without the
usual shortcomings such as the caustics and non-causal
implications. It turned out that the backreaction in the gauge
chosen mimics a normal matter with positive pressure, and a
density roughly 4 to 5 order of magnitudes less than the matter
density. This is in contrast to vanishing of backreaction in other
gauges, and confirm the claims that backreaction is gauge
dependent.
 We therefore conclude that, although the method can be
made free of singularities and superluminal effects, the volume
averaging in the SFRW toy model leads to a non-vanishing addition
to the normal matter five order of magnitudes less than the matter
density, indicating a kind of dark matter component, but no sign
of any dark energy component.

\bibliographystyle{unsrt}
\bibliography{Average}

\begin{thebibliography}{10}

\bibitem{Mansouri05}
R.~{Mansouri}.
\newblock {Structured FRW universe leads to acceleration: a non-perturbative
  approach}.
\newblock {\em arxiv:astro-ph/0512605}, December 2005.

\bibitem{Celer07}
M.-N. {C{\'e}l{\'e}rier}.
\newblock {The accelerated expansion of the Universe challenged by an effect of
  the inhomogeneities. A review}.
\newblock {\em arxiv:astro-ph/0702416}, February 2007.

\bibitem{Chuang}
C.-H. {Chuang}, J.-A. {Gu}, and W.~P. {Hwang}.
\newblock {Inhomogeneity-Induced Cosmic Acceleration in a Dust Universe}.
\newblock {\em arxiv:astro-ph/0512651}, December 2005.

\bibitem{Kolb07}
V.~{Marra}, E.~W. {Kolb}, S.~{Matarrese}, and A.~{Riotto}.
\newblock {On cosmological observables in a swiss-cheese universe}.
\newblock {\em arXiv:0708.3622}, August 2007.

\bibitem{Ellis84}
G.~F.~R. {Ellis}.
\newblock {General relativity and gravitation}.
\newblock In B.~{Bertotti}, F.~{de Felice}, and A.~{Pascolini}, editors, {\em
  General Relativity and Gravitation Conference}, page 215, 1984.

\bibitem{Ellis87}
G.~F.~R. {Ellis} and W.~{Stoeger}.
\newblock {The 'fitting problem' in cosmology}.
\newblock {\em Classical and Quantum Gravity}, 4:1697--1729, November 1987.

\bibitem{Bild91}
S.~{Bildhauer} and T.~{Futamase}.
\newblock {The cosmic microwave background in a globally inhomogeneous
  universe}.
\newblock {\em \mnras}, 249:126--130, March 1991.

\bibitem{Futa93}
T.~{Futamase}.
\newblock {General Relativistic Description of a Realistic Inhomogeneous
  Universe}.
\newblock {\em Progress of Theoretical Physics}, 89:581--597, March 1993.

\bibitem{Kolb05}
E.~W. {Kolb}, S.~{Matarrese}, and A.~{Riotto}.
\newblock {On cosmic acceleration without dark energy}.
\newblock {\em arxiv:astro-ph/0506534}, June 2005.

\bibitem{Romano}
A.~{Enea Romano}.
\newblock {LTB universes as alternatives to dark energy: does positive averaged
  acceleration imply positive cosmic acceleration?}
\newblock {\em arxiv:astro-ph/0612002}, November 2006.

\bibitem{Singh1}
A.~{Paranjape} and T.~P. {Singh}.
\newblock {The possibility of cosmic acceleration via spatial averaging in
  Lema{\^i}tre Tolman Bondi models}.
\newblock {\em Classical and Quantum Gravity}, 23:6955--6969, December 2006.

\bibitem{Singh2}
A.~{Paranjape} and T.~P. {Singh}.
\newblock {Explicit Cosmological Coarse Graining via Spatial Averaging}.
\newblock {\em arxiv:astro-ph/0609481}, September 2006.

\bibitem{Notari05}
E.~W. {Kolb}, S.~{Matarrese}, A.~{Notari}, and A.~{Riotto}.
\newblock {Primordial inflation explains why the universe is accelerating
  today}.
\newblock {\em hep-th/0503117}, March 2005.

\bibitem{Zalal92}
R.~M. {Zalaletdinov}.
\newblock {Averaging out the Einstein equations}.
\newblock {\em General Relativity and Gravitation}, October 1992.

\bibitem{Zalal93}
R.~M. {Zalaletdinov}.
\newblock {Towards a theory of macroscopic gravity}.
\newblock {\em General Relativity and Gravitation}, July 1993.

\bibitem{Buchert00}
T.~{Buchert}.
\newblock {On Average Properties of Inhomogeneous Fluids in General Relativity:
  Dust Cosmologies}.
\newblock {\em General Relativity and Gravitation}, 32:105--126, January 2000.

\bibitem{Buchert01}
T.~{Buchert}.
\newblock {On Average Properties of Inhomogeneous Fluids in General Relativity:
  Perfect Fluid Cosmologies}.
\newblock {\em General Relativity and Gravitation}, 33:1381--1405, August 2001.

\bibitem{Buch-Ehl-1}
T.~{Buchert} and J.~{Ehlers}.
\newblock {Averaging inhomogeneous Newtonian cosmologies.}
\newblock {\em \aap}, 320:1--7, April 1997.

\bibitem{Ishi06}
A.~{Ishibashi} and R.~M. {Wald}.
\newblock {Can the acceleration of our universe be explained by the effects of
  inhomogeneities?}
\newblock {\em Classical and Quantum Gravity}, 23:235--250, January 2006.

\bibitem{Khosr07}
S.~{Khosravi}, E.~{Kourkchi}, R.~{Mansouri}, and Y.~{Akrami}.
\newblock {Structured FRW universe based on LTB junctions along the past light
  cone}.
\newblock {\em arxiv:astro-ph/0702282}, February 2007.

\bibitem{khakshour07}
S.~{Khakshournia} and R.~{Mansouri}.
\newblock {Matching LTB and FRW spacetimes through a null hypersurface}.
\newblock {\em gr-qc/0702130}, February 2007.

\bibitem{Vander06}
R.~A. {Vanderveld}, {\'E}.~{\'E}. {Flanagan}, and I.~{Wasserman}.
\newblock {Mimicking dark energy with Lema{\^i}tre-Tolman-Bondi models: Weak
  central singularities and critical points}.
\newblock {\em \prd}, 74(2):023506--+, July 2006.

\bibitem{Newman86}
R.~P.~A.~C. {Newman}.
\newblock {Strengths of naked singularities in Tolman-Bondi spacetimes}.
\newblock {\em Classical and Quantum Gravity}, 3:527--539, July 1986.

\bibitem{Behrend}
J.~{Behrend}, I.~A. {Brown}, and G.~{Robbers}.
\newblock {Cosmological Backreaction from Perturbations}.
\newblock {\em arxiv:0710.4964}, October 2007.

\bibitem{Stewart}
J.~M. {Stewart} and M.~{Walker}.
\newblock {Perturbations of space-times in general relativity}.
\newblock {\em Royal Society of London Proceedings Series A}, 341:49--74,
  October 1974.

\bibitem{Schwarz}
N.~{Li} and D.~J. {Schwarz}.
\newblock {On the onset of cosmological backreaction}.
\newblock {\em gr-qc/0702043}, February 2007.

\bibitem{Ghassemi}
S.~{Ghassemi}, S.~{Khoeini-Moghaddam}, and R.~{Mansouri}.
\newblock In preparation.

\bibitem{Mattsson}
T.~{Mattsson} and M.~{Ronkainen}.
\newblock {Exploiting scale dependence in cosmological averaging}.
\newblock {\em arXiv:0708.3673}, August 2007.

\end{thebibliography}

\end{document}